\documentclass[aip,longbiblography
]{revtex4-1}
\usepackage{chemformula}
\usepackage{graphicx}
\usepackage{amsmath,amsfonts,amssymb}
\usepackage{siunitx}
\draft 
\usepackage{mathrsfs}
\usepackage[scr=rsfs,cal=boondox]{mathalfa}
\begin{document}


\title[Molecular Beam Epitaxy of \ch{AgTaO3}]{Molecular Beam Epitaxy of \ch{AgTaO3}} 



\author{Tobias Schwaigert}
\affiliation{Platform for the Accelerated Realization, Analysis, and Discovery of Interface Materials (PARADIM), Cornell University, Ithaca, New York 14853, USA}
\affiliation{Department of Materials Science and Engineering, Cornell University, Ithaca, NY 14853, USA}
\author{Joshua Maile}
\affiliation{Department of Materials Science and Engineering, Ohio State University, Columbus, OH 43210 USA}
\author{Olivia Peek}
\affiliation{Platform for the Accelerated Realization, Analysis, and Discovery of Interface Materials (PARADIM), Cornell University, Ithaca, New York 14853, USA}
\affiliation{Department of Physics Cornell University, Ithaca, NY 14853, USA}
\author{Eric Biedke}
\affiliation{Department of Materials Science and Engineering, Ohio State University, Columbus, OH 43210 USA}
\author{Paul T. Malinowski}
\affiliation{Department of Physics Cornell University, Ithaca, NY 14853, USA}
\author{Kyle M. Shen}
\affiliation{Department of Physics Cornell University, Ithaca, NY 14853, USA}
\affiliation{Laboratory of Atomic and Solid State Physics, Department of Physics, Cornell University, Ithaca, NY, USA}
\author{Salva Salmani-Rezaie}
\affiliation{Department of Materials Science and Engineering, Ohio State University, Columbus, OH 43210 USA}
 \author{Darrell G. Schlom}
\affiliation{Platform for the Accelerated Realization, Analysis, and Discovery of Interface Materials (PARADIM), Cornell University, Ithaca, New York 14853, USA}
\affiliation{Department of Materials Science and Engineering, Cornell University, Ithaca, NY 14853, USA}
\affiliation{Kavli Institute at Cornell for Nanoscale Science, Cornell University, Ithaca, New York 14853, USA}
\affiliation{Leibniz-Institut für Kristallzüchtung, Max-Born-Str. 2, 12489 Berlin, Germany}
\author{Kaveh Ahadi}
\email[ahadi.4@osu.edu]{}
\affiliation{Department of Materials Science and Engineering, Ohio State University, Columbus, OH 43210 USA}
\affiliation{Department of Electrical and Computer Engineering, Ohio State University, Columbus, OH 43210 USA}

\date{\today}

\begin{abstract}

We report the first synthesis of single-crystal \ch{AgTaO3} thin films using molecular-beam epitaxy (MBE). High-quality epitaxial \ch{AgTaO3} films were grown on both (001)- and (111)-oriented \ch{SrTiO3} substrates using sequential deposition of atomic silver and \ch{TaO2} layers under an ozone/oxygen atmosphere (80 \% \ch{O3} + 20 \% \ch{O2}). X-ray diffraction and reciprocal space mapping demonstrate that the films are coherently strained to the \ch{SrTiO3} substrates with sharp rocking curves comparable to the substrates, indicating high structural perfection. High-angle annular dark-field scanning transmission electron microscopy (HAADF-STEM) confirms coherent, low defect growth for $(001)_{pc}$-oriented films. For $(111)_{pc}$-oriented films, initial coherent growth proceeds up to ~10 nm before transitioning into a Ta-rich surface region. Energy-dispersive X-ray spectroscopy (EDX) reveals a narrow cation intermixing region at the substrate interface for both orientations. This work demonstrates an effective synthesis route for single-crystal \ch{AgTaO3} thin films, providing a platform to investigate strain engineering and emergent interfacial properties in silver-based tantalates. 

\end{abstract}


\maketitle 

\section{Introduction\label{sec:level1}}
Tantalate perovskites have emerged as a promising material platform to explore quantum phenomena in bulk and at oxide interfaces, driven by the combination of strong $5d$ spin-orbit coupling and enhanced permittivity and carrier mobility.\cite{gupta2022ktao3, schwaigert2026synthesis, al2023enhanced} When integrated into low-dimensional heterostructures, these features could give rise to two-dimensional electron gases\cite{al2021two, zou2015latio3} with robust interfacial superconductivity\cite{liu2021two, arnault2023anisotropic, al2022superconductivity} and efficient spin-charge interconversion.\cite{al2025spin, vicente2021spin} Realizing the full potential of these emergent properties, however, hinges on the ability to synthesize single-crystal thin films with precise stoichiometry, atomic-scale interface sharpness, and minimal defect densities.\cite{kim2024electronic} Despite recent interest in quantum phases in \ch{KTaO3}, \cite{al2022oxygen, mccourt2025electrostatic, poage2025violation} less attention has been given to the \ch{Ag^+} analog. \ch{AgTaO3} exhibits relatively high permittivity, low loss tangent and potential as an anti-ferroelectric storage material \cite{zhao2017lead,valant2007review,liu2018antiferroelectrics}.

First synthesized by Francombe and Lewis in 1958,\cite{francombe1958structural} the room-temperature crystal structure of \ch{AgTaO3} has been a subject of debate. Initially, the room temperature structure was believed to be orthorhombic.\cite{francombe1958structural} Later, Belyaev assigned a polar rhombohedral structure ($R3c$) to room temperature \ch{AgTaO3}.\cite{belyaev1978orthorhombic} A permittivity anomaly typically marks the polar transition. Soon et al. \cite{soon2010dielectric} did not observe an anomaly in permittivity and proposed a centro-symmetric rhombohedral structure ($R\overline{3}m$) for room-temperature \ch{AgTaO3}, while others reported such anomaly.\cite{suchanicz2009uniaxial} More recent diffraction results suggest the following sequence of phase transitions for bulk \ch{AgTaO3}: $R3c$\ch{<->[400 °C]}$Ibmm$\ch{<->[430 °C]}$P4/mbm$\ch{<->[500 °C]}$Pm\overline{3}m$.\cite{li2025high} Furthermore, epitaxial synthesis parameters such as epitaxial strain and substrate clamping effects could impact the structural phase diagrams in epitaxial thin layers.\cite{PhysRevB.69.212101, PhysRevLett.80.1988, schlom2007strain}

In conventional solid-state synthesis, \ch{AgTaO3} is formed by reacting \ch{Ag2O} and \ch{Ta2O5} powders. Thermo-gravimetric analysis performed by Valant et al. showed that \ch{Ag2O} decomposes to metallic silver before the perovskite formation, suggesting \ch{Ag^+} is only recovered once silver reacts with \ch{Ta2O5} and oxygen at higher temperatures.\cite{valant2007review} Thin-film deposition techniques that supply individual atomic species could circumvent this thermodynamic barrier. Despite the successful bulk synthesis of \ch{AgTaO3}, its thin film synthesis has remained limited \cite{tachikawa2017enhancing} and of this materials system by molecular-beam epitaxy (MBE) has not been demonstrated. A synthesis approach for high-quality single crystalline thin films could resolve competing structures \cite{khan2021structural} and provide a platform to investigate the impact of epitaxial tuning. Furthermore, recent molecular-beam epitaxy of close relative \ch{KTaO3}\cite{schwaigert2023molecular} has enabled the epitaxial strain stabilization of a robust ferroelectric state above room temperature.\cite{schwaigert2026above} 

In this work, we present the first MBE synthesis of \ch{AgTaO3}. \ch{AgTaO3} films are grown on \ch{SrTiO3} (100) and \ch{SrTiO3} (111) single crystal substrates. X-ray diffraction reveals that the films are coherently strained and have high quality, comparable to \ch{SrTiO3} single crystal substrate. HAADF-STEM images confirm that the films have abrupt interfaces and do not show any extended defects.

\section{Experimental}
Epitaxial \ch{AgTaO3} films were grown using a modified Vecco Gen 10 MBE system, where the conventional SiC heater was replaced by 10 \unit{\um} \ch{CO2}-laser from Epiray GmbH (THERMALAS Substrate heater). A molecular beam of \ch{TaO2} was generated from an effusion cell containing \ch{Ta2O5} (Alfa Aesar, 99.993 \%). \cite{adkison2020, yorick2026} A molecular beam of silver was generated by a conventional effusion cell  (Alfa Aesar, 99.999 \%). Films were grown by sequential deposition of one silver atomic layer followed by one atomic layer of \ch{TaO2} in a substrate temperature range of 500-600 °C, measured by an optical pyrometer operating at 7.5 \unit{\um} wavelength. The silver flux was measured before growth using a quartz crystal microbalance (1-2 $\times 10^{13}$ atoms/$\text{cm}^2$/s) while the \ch{TaO2} was determined by deposition on R-plane sapphire. The deposition rate was confirmed by x-ray reflectometry measurements. A mixture of ozone and oxygen (80 \% \ch{O3} + 20 \% \ch{O2}) was used as the oxidant. The films were grown at an oxidant background pressure of $1\times10^{-5}$ Torr. The (001) and (111) \ch{SrTiO3} substrates were used as received. 

X-ray diffraction (XRD), X-ray reflectometry (XRR), and reciprocal space mapping (RSM) measurements were carried out using a PANalytical Empyrean diffractometer with Cu K$\alpha_{1}$ radiation. The raw XRR spectra were analyzed using the PANalytical X´Pert Reflectivity software package and the layer thickness was derived from a fast Fourier transform (FFT) after manually defining the critical angle to account for refractive effects. \textit{In situ} X-ray photoelectron spectroscopy (XPS) was performed in the same chamber using a non-monochromated Scienta Omicron DSX400 x-ray source. Spectra were analyzed with an Omicron Sphera II analyzer after excitation. \textit{In situ} reflection high-energy electron diffraction (RHEED) patterns were recorded using KSA-400 software and a Staib electron source operated at  14 kV and a filament current of 1.5 A. The morphology of the film surface was characterized using an Asylum Cypher ES environmental AFM. Cross-sectional TEM specimens were prepared by focused ion beam milling. Initial milling was performed at 30 kV, followed by final thinning and polishing at 5 kV to reduce ion-beam-induced surface damage. STEM imaging was carried out using a Thermo Fisher Scientific Themis Z operated at 200 kV with a probe semi-convergence angle of 20 mrad. High-angle annular dark-field STEM (HAADF-STEM) images were acquired over a detector collection range of 64–200 mrad. To improve the signal-to-noise ratio while minimizing the effects of sample drift, 20 fast-scan images, each containing 2048 × 2048 pixels with a dwell time of 200 ns per pixel, were aligned and averaged. Energy-dispersive X-ray spectroscopy (EDX) spectrum imaging was performed using a Super-X detector, and elemental maps were generated from the net X-ray counts.

\section{Results}

\begin{figure}[p]
\includegraphics[width=0.8\textwidth]{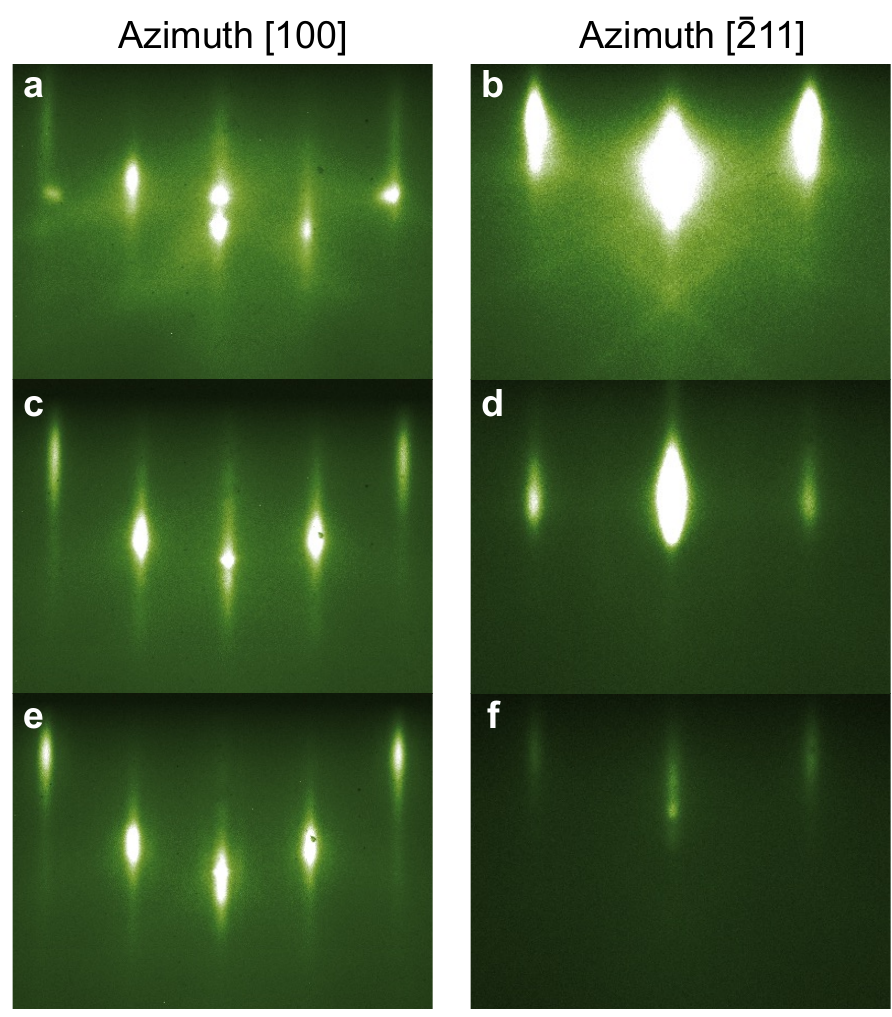}
\caption{\label{fig:RHEED} RHEED pattern after growth of a formula-unit-thick layer \ch{AgTaO3} on (a) (001) \ch{SrTiO3} and (b) (111) \ch{SrTiO3} substrates. RHEED pattern after growth of 20 formula-unit-thick layers of \ch{AgTaO3} on (c) (001) \ch{SrTiO3} and (d) (111) \ch{SrTiO3} substrates. RHEED pattern immediately after the growth of (e) 
28 nm \ch{AgTaO3} on (001) \ch{SrTiO3} and (f) 17 nm \ch{AgTaO3} on (111) \ch{SrTiO3} substrates.} 
\end{figure}

The bulk \ch{AgTaO3}  lattice constants are $a=b=5.528$~\AA~and $c=13.715$~\AA~ with rhombohedral space group symmetry (\textit{R}3\textit{c}, $a^-a^-a^-$ Glazer notation) at room temperature.\cite{wolcyrz1986} Accordingly, \ch{AgTaO3} grows pseudocube-on-cube on \ch{SrTiO3}. The expected crystallographic orientation of \ch{AgTaO3} grown on \ch{SrTiO3} (001) and (111) are (012)$_R$ and (001)$_R$, receptively, corresponding to the (001)$_{pc}$ and (111)$_{pc}$. Here the subscript \textit{R} denotes rhombohedral indices and \textit{pc} denotes pseudocubic indices.  The lattice mismatch between 
(001)$_{pc}$ \ch{AgTaO3}  ($a_{pc}=3.9258$ ~\AA) and (001) \ch{SrTiO3} ($a_{STO}=3.9051$ ~\AA) is $-0.53$\%.  The lattice mismatch between  (111)$_{pc}$ \ch{AgTaO3}  ($a_{pc}=5.5281$ ~\AA) and (111) \ch{SrTiO3} ($a_{STO}=5.5226$ ~\AA) is $-0.1$\%. 
Reflection high-energy electron diffraction (RHEED) was used to monitor the evolution of the surface structure and reconstruction during growth of both epitaxial orientations. Figures \ref{fig:RHEED}(a) and \ref{fig:RHEED}(b) show the RHEED diffraction pattern after the growth of first formula-unit-thick of (001)$_{pc}$ and (111)$_{pc}$ \ch{AgTaO3} films on (001) and (111) \ch{SrTiO3}, respectively. Along their high symmetry directions diffraction streaks and Kikuchi lines are visible. We note the small additional dots at (001)$_{pc}$ \ch{AgTaO3} orientation. Figures \ref{fig:RHEED}(c) and (d) show the RHEED pattern after the growth of a 20 formula-unit-thick layer of (001)$_{pc}$ and (111)$_{pc}$ \ch{AgTaO3}, respectively. The additional dot-like features observed in the (001)$_{pc}$ orientation have vanished and clear diffraction pattern of \ch{AgTaO3} is visible. Figures \ref{fig:RHEED}(e) and (f) exhibit the RHEED pattern immediately after the growth of 28 nm thick (001)$_{pc}$ and 17 nm thick (111)$_{pc}$ \ch{AgTaO3} films, where the shutters of both silver and \ch{TaO2} are closed, but the substrate is still immersed in ozone and at growth temperature. While the (001)$_{pc}$ \ch{AgTaO3} RHEED pattern remains clear, the (111)$_{pc}$ \ch{AgTaO3} appears cloudy.

\textit{In situ} X-ray photoelectron spectroscopy (XPS) was used to investigate the oxidation states of the\ch{AgTaO3} thin films (Fig. S1). The Ag $3d$ spectrum exhibits a spin-orbit doublet with the 3$d_{5/2}$ and 3$d_{3/2}$ components centered at 368.10 eV (FWHM = 1.58 eV) and 374.12 eV (FWHM = 1.55 eV), respectively, corresponding to a spin-orbit splitting of 6.01 eV and an intensity ratio of approximately 1.46:1 (close to the statistical 3:2 ratio). Assigning this binding energy to a specific silver oxidation state is complicated by substantial overlap reported in the literature.\cite{kaspar2010spectroscopic} The Ta $4f$ doublet at 4$f_{7/2}$= 26.25 eV and 4$f_{5/2}$= 28.12 eV is consistent with the \ch{Ta^{5+}} oxidation state.\cite{schwaigert2026synthesis,KHANUJA200941}

\begin{figure}[p]
\includegraphics[width=\textwidth]{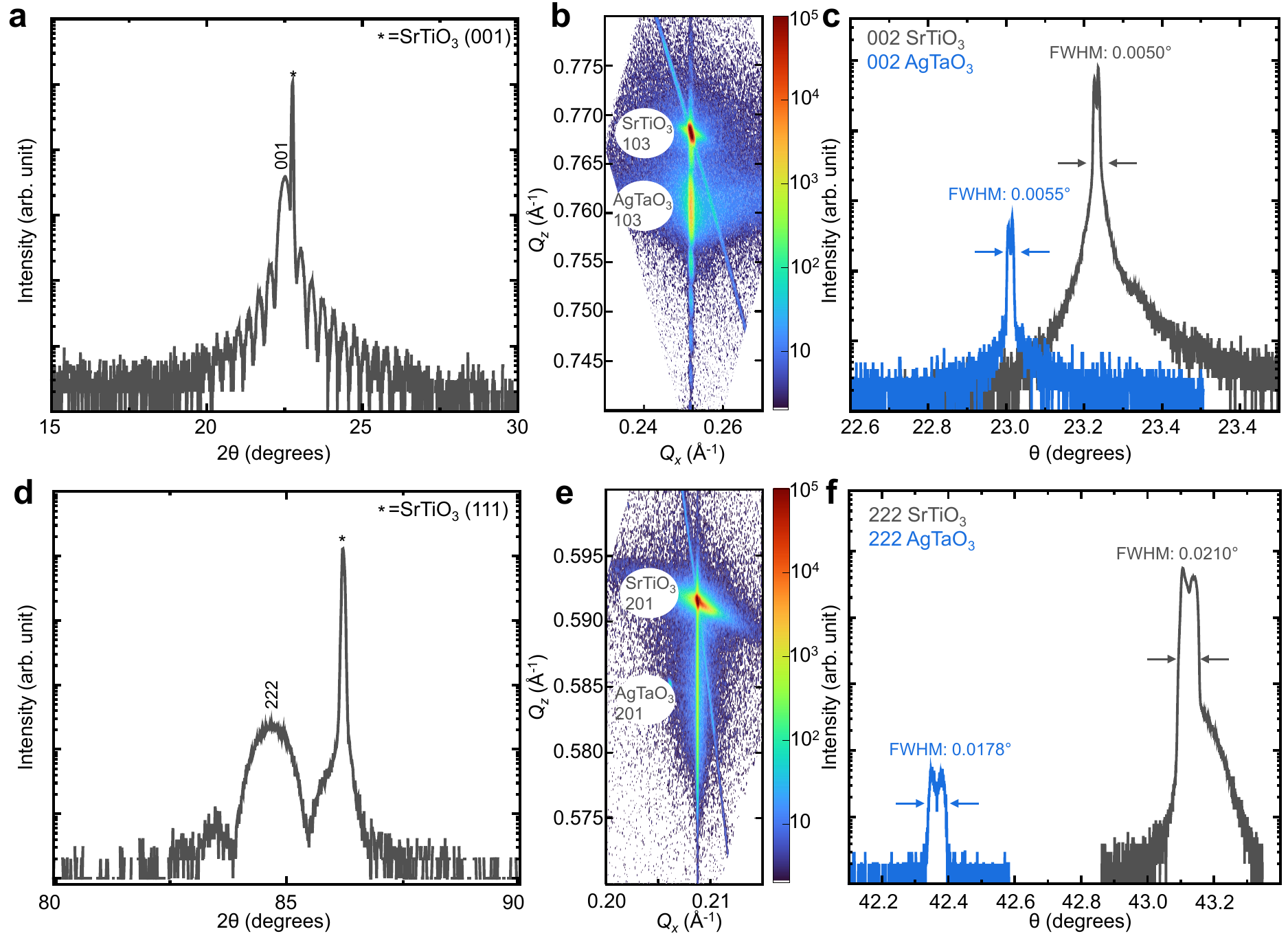}
\caption{\label{fig:XRD} (a) $\theta$-2$\theta$ scan, showing 001 reflection of \ch{AgTaO3} grown on \ch{SrTiO3} (001) substrate. Symmetric Laue fringes indicate a well-defined film thickness, indicative of an abrupt interface between film and substrate (asterisks * denote substrate reflections). (b) RSM around the 103 substrate and film reflections. The RSM results confirm that the film is fully strained to the substrate. (c) Overlaid rocking curves of the  002 \ch{SrTiO3} and 002 \ch{AgTaO3} reflections suggest comparable FWHMs, indicating low out-of-plane mosaicity. (d) $\theta$-2$\theta$ scan, showing 222 reflection of \ch{AgTaO3} grown on a \ch{SrTiO3} (111) substrate (asterisks * denote substrate reflections). (b) RSM around the 201 substrate reflection. The RSM results confirm that the film is coherently strained to the substrate. (c) Overlaid rocking curves of the 222 \ch{SrTiO3} and 222 \ch{AgTaO3} reflections exhibit comparable FWHMs, suggesting low out-of-plane mosaicity.} 
\end{figure}

Atomic force microscopy (AFM) was used to investigate the surface \textit{ex situ} (Figs. S2 and S3). The root-mean-square roughness, measured by taking a \SI{1}{\um^2} area as a reference, is 0.94 nm and 0.82 nm for the (001)$_{pc}$ and (111)$_{pc}$ oriented \ch{AgTaO3} films, respectively. Figure \ref{fig:XRD} shows the x-ray diffraction results of the same 28 nm  (001)$_{pc}$ and 17 nm (111)$_{pc}$ thick \ch{AgTaO3} films. Figure \ref{fig:XRD}(a) shows the $\theta$-2$\theta$ scan around 001 \ch{AgTaO3} reflection, surrounded by Laue fringes.\cite{friedrich1913} The full spectra can be found in the supplemental material (Fig. S4 and S5). An RSM around the 103 substrate reflection confirms that the (001)$_{pc}$ \ch{AgTaO3} film is coherently strained to the substrate. Figure \ref{fig:XRD}(c) shows the overlaid rocking curves of the 002 \ch{AgTaO3} and \ch{SrTiO3} peaks. The full width at half maximum (FWHM) for both the film (about 20 arcsec) and substrate (about 18 arcsec) are comparable and suggest high crystalline quality of the film. Figure \ref{fig:XRD}(d) shows the 222 peak of 17 nm thick \ch{AgTaO3} film grown on a \ch{SrTiO3} (111) substrate. An RSM around the 201 substrate peak confirms that the film is coherently strained to the substrate. The FWHM for both the film (about 64 arcsec) and substrate (about 76 arcsec) are comparable.  

Using the full $\theta$-2$\theta$ scan and following the Nelson-Riley procedure,\cite{nelson1945} for the film grown on \ch{SrTiO3}(001), the out-of-plane pseudocubic lattice constant is resolved $a_{pc}$ = 3.9429 \AA. Our out-of-plane lattice constant is larger than the \ch{AgTaO3} film with similar thickness grown on \ch{SrTiO3}(001) by pulsed-laser depostion (3.935 \AA).\cite{tachikawa2017enhancing}  A comparable discrepancy has recently been reported between the measured and expected lattice parameters for \ch{KTaO3} films grown on \ch{SrTiO3}, where lattice expansion was attributed to the emergence of a ferroelectric state.\cite{schwaigert2026above} The unequivocal resolution of the space group symmetries and the nature of expanded out-of-plane lattice constant require complementary characterizations.
For the (111)$_{pc}$ \ch{AgTaO3} the extracted  lattice spacing $a_{ATO_{111}} $  is 2.283 \AA, ~ which is comparable to the expected 2.285 \AA ~of the unperturbed rhombohedral crystal structure. This close agreement is likely due to the small lattice mismatch with (111) \ch{SrTiO3} ( $-0.1$\%).

\begin{figure}[p]
\includegraphics[width=\textwidth]{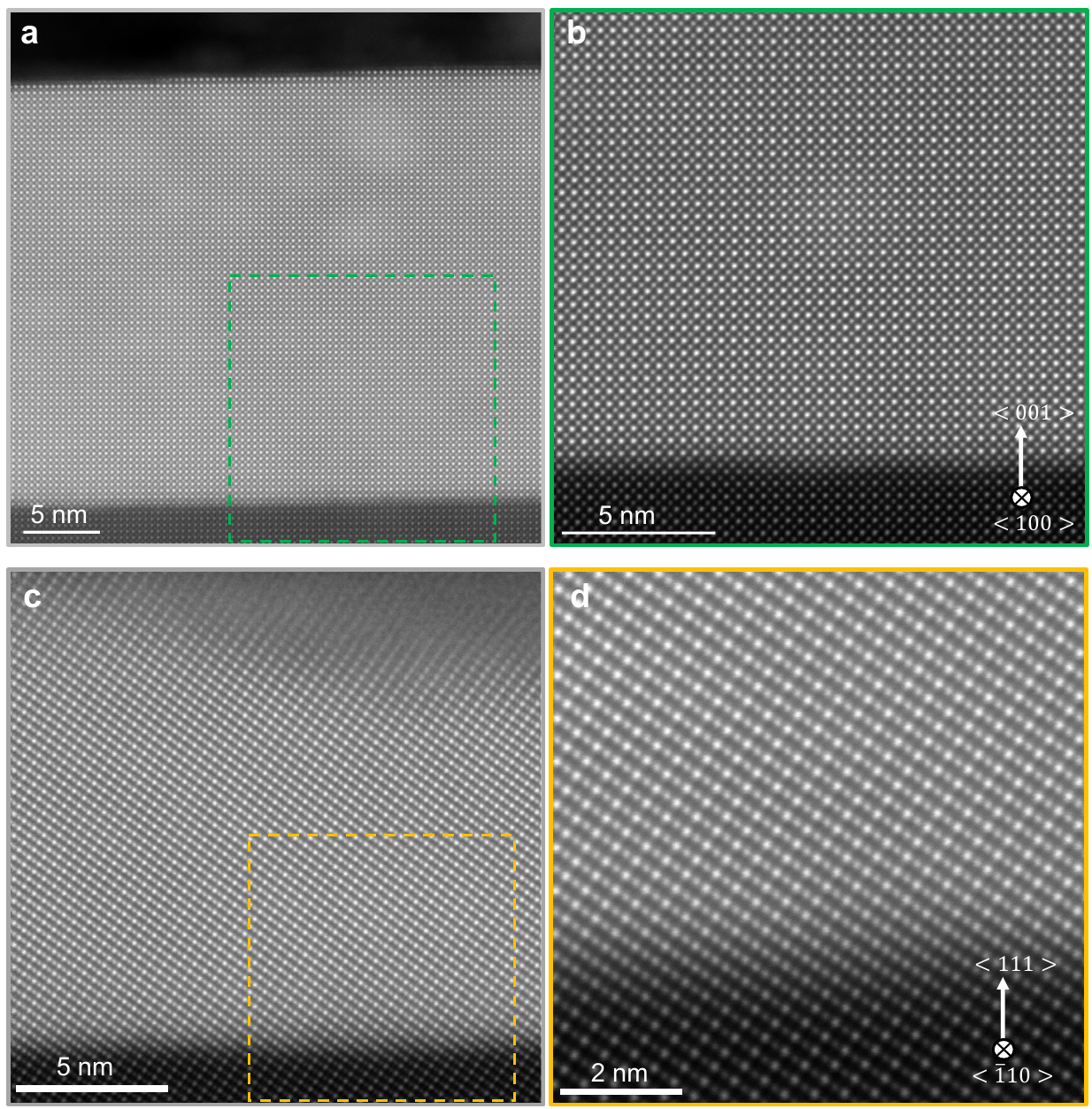}
\caption{\label{fig:STEM001}Cross-section HAADF-STEM images of the same 28 nm thick (001) and 17 nm (111) \ch{AgTaO3} films. (a)  At low magnification surface steps are visible with and extended regions of uniformity for the (001) \ch{AgTaO3} film. (b) Higher magnification  reveal well-resolved atomic columns within the film and a sharp interface between the film and underlying (001) \ch{SrTiO3} substrate. (c) (111) \ch{AgTaO3} at low magnification reveals  $\approx$ 10 nm of well-ordered crystalline film, above which the structure transitions into a tantalum-rich oxide layer. (d) Higher magnification reveals the sharp film-substrate interface.}
\end{figure}

Imaging by high-angle annular dark-field in a scanning transmission electron microscope (HAADF-STEM) was used to further investigate the \ch{AgTaO3} films. The (001)$_{pc}$ \ch{AgTaO3} sample shows a rather featureless image, suggesting a high crystalline quality (Fig. \ref{fig:STEM001}). We do not observe any extended defects and the interface appears coherent. We also used EDX elemental mapping to characterize the grown films (Fig. S6). EDX elemental maps and the corresponding intensity profiles show a familiar narrow transition region in which the silver and tantalum signals decrease while the strontium and titanium signals emerge, indicating cation intermixing across the \ch{AgTaO3}/\ch{SrTiO3} interface. Although a two-dimensional electron gas can accommodate the electrostatic discontinuity at the interface,\cite{ahadi2017novel, mori2019controlling} atomic intermixing may play a similar role.\cite{schwaigert2023molecular}

The (111)$_{pc}$ \ch{AgTaO3} film exhibits a high crystalline quality for the first 10 nm from the substrate interface followed by a gradually increasing disorder into a tantalum-rich oxide layer near the surface. EDX profiles display intermixing at the film and substrate interface similar to the (001)$_{pc}$ \ch{AgTaO3} sample. The formation of the tantalum-rich disordered surface region could be associated with compensation of the highly polar (111) surface, potentially coupled with cation redistribution or silver loss during growth. EDX profiles also confirm progressive silver depletion toward the film surface, while the tantalum and oxygen signals persist (Fig. S7), consistent with the formation of the tantalum-rich surface region. Because the unreconstructed \ch{AgTaO3}(111)$_{pc}$ surface is itself polar, surface reconstruction and cation redistribution may contribute to the formation of this region. Finally, other growth-related mechanisms, including preferential silver loss, may also contribute to the tantalum-rich surface region observed.

In summary, we demonstrate MBE growth of high-quality \ch{AgTaO3} for both (001)$_{pc}$ and (111)$_{pc}$ oriented films. XRD results suggest high crystalline perfection, and RSM confirms that the films are coherently strained to the underlying substrates. Cross-sectional HAADF-STEM confirms the high crystalline quality of the samples and does not show any extended defects. EDX results show that in both orientations, the transition region between the substrate and film is not atomically abrupt.

 
 \begin{acknowledgments}
This material is based upon work supported by the National Science Foundation (Platform for the Accelerated Realization, Analysis, and Discovery of Interface Materials (PARADIM)) under Cooperative Agreement No. DMR-2039380. E.B is supported by the U.S. National Science Foundation under Grant No. NSF DMR-2408890. 

\end{acknowledgments}
\appendix
\section*{AUTHOR DECLARATIONS}
\textbf{Conflict of Interest}

The authors have no conflicts to disclose.

\section*{Data Availability Statement}
The data that support the findings of this study are available within the article. \textbf{Additional data related to the} film growth and structural characterization by XRD and STEM \textbf{are available at https://doi.org/10.34863/xxxx}

\bibliography{aipsamp.bib}

@PREAMBLE{
 "\providecommand{\noopsort}[1]{}" 
 # "\providecommand{\singleletter}[1]{#1}%" 
}

@article{adkison2020,
  title={{Suitability of binary oxides for molecular-beam epitaxy source materials: A comprehensive thermodynamic analysis}},
  author={Adkison, Kate M and Shang, Shun-Li and Bocklund, Brandon J and Klimm, Detlef and Schlom, Darrell G and Liu, Zi-Kui},
  journal={APL Mater.},
  volume={8},
  number={8},
  pages={081110},
  year={2020},
  publisher={AIP Publishing LLC}
}

@article{al2021two,
  title={{Two-dimensional electron systems and interfacial coupling in \ch{LaCrO3}/\ch{KTaO3} heterostructures}},
  author={Al-Tawhid, Athby H and Kumah, Divine P and Ahadi, Kaveh},
  journal={Appl. Phys. Lett.},
  volume={118},
  number={19},
  pages={192905},
  year={2021},
  publisher={AIP Publishing LLC}
}

@article{al2022oxygen,
  title={{Oxygen Vacancy-Induced Anomalous Hall Effect in a Nominally Non-magnetic Oxide}},
  author={Al-Tawhid, Athby H and Kanter, Jesse and Hatefipour, Mehdi and Irving, Douglas L and Kumah, Divine P and Shabani, Javad and Ahadi, Kaveh},
  journal={J. Electron. Mater.},
  volume={51},
  number={12},
  pages={7073--7077},
  year={2022},
  publisher={Springer}
}

@article{al2022superconductivity,
  title={Superconductivity and Weak Anti-localization at KTaO3 (111) Interfaces: AH Al-Tawhid et al.},
  author={Al-Tawhid, Athby H and Kanter, Jesse and Hatefipour, Mehdi and Kumah, Divine P and Shabani, Javad and Ahadi, Kaveh},
  journal={Journal of Electronic Materials},
  volume={51},
  number={11},
  pages={6305--6309},
  year={2022},
  publisher={Springer}
}

@article{mori2019controlling,
  title={Controlling a Van Hove singularity and Fermi surface topology at a complex oxide heterostructure interface},
  author={Mori, Ryo and Marshall, Patrick B and Ahadi, Kaveh and Denlinger, Jonathan D and Stemmer, Susanne and Lanzara, Alessandra},
  journal={Nature communications},
  volume={10},
  number={1},
  pages={5534},
  year={2019},
  publisher={Nature Publishing Group UK London}
}

@article{ahadi2017novel,
  title={Novel metal-insulator transition at the SmTiO 3/SrTiO 3 interface},
  author={Ahadi, Kaveh and Stemmer, Susanne},
  journal={Physical Review Letters},
  volume={118},
  number={23},
  pages={236803},
  year={2017},
  publisher={APS}
}

@article{wolcyrz1986,
title = {{The crystal structure of the room-temperature phase of \ch{AgTaO3}}},
author={Wolcyrz, M and  Lukaszewski, M},
pages = {53--58},
volume = {177},
number = {1-2},
journal = {Z. Kristallogr. - Cryst. Mater.},
doi = {doi:10.1524/zkri.1986.177.1-2.53},
year = {1986},
}

@article{tachikawa2017enhancing,
  title={{Enhancing the barrier height in oxide Schottky junctions using interface dipoles}},
  author={Tachikawa, Takashi and Hwang, Harold Y and Hikita, Yasuyuki},
  journal={Appl. Phys. Lett.},
  volume={111},
  number={9},
  year={2017},
  publisher={AIP Publishing}
}

@article{schwaigert2026above,
  title={Above Room Temperature Ferroelectricity in Epitaxially Strained KTaO3},
  author={Schwaigert, Tobias and Salmani-Rezaie, Salva and Hazra, Sankalpa and Saha, Utkarsh and Ramesh, Maya and Ross, Aiden and Pamuk, Bet{\"u}l and Chen, Long-Qing and Muller, David A and Schlom, Darrell G and Ahadi, Kaveh},
  journal={Advanced Materials},
  pages={e74366},
  year={2026},
  publisher={Wiley Online Library}
}

@article{valant2007review,
  title={{Review of \ch{Ag(Nb,Ta)O3} as a functional material}},
  author={Valant, Matjaz and Axelsson, Anna-Karin and Alford, Neil},
  journal={J. Eur. Ceram. Soc.},
  volume={27},
  number={7},
  pages={2549--2560},
  year={2007},
  publisher={Elsevier}
}

@article{liu2021two,
  title={{Two-dimensional superconductivity and anisotropic transport at \ch{KTaO3} (111) interfaces}},
  author={Changjiang Liu  and Xi Yan  and Dafei Jin  and Yang Ma  and Haw-Wen Hsiao  and Yulin Lin  and Terence M. Bretz-Sullivan  and Xianjing Zhou  and John Pearson  and Brandon Fisher  and J. Samuel Jiang  and Wei Han  and Jian-Min Zuo  and Jianguo Wen  and Dillon D. Fong  and Jirong Sun  and Hua Zhou  and Anand Bhattacharya },
  journal={Science},
  volume={371},
  number={6530},
  pages={716--721},
  year={2021},
  publisher={American Association for the Advancement of Science}
}

@article{yorick2026,
    author = {Birkhölzer, Yorick A. and Park, Anna S. and Schnitzer, Noah and Kaaret, Jeffrey Z. and Gregory, Benjamin Z. and Kraay, Tomas A. and Schwaigert, Tobias and Barone, Matthew R. and Faeth, Brendan D. and Hensling, Felix V. E. and van den Bosch, Iris C. G. and Kiens, Ellen M. and Baeumer, Christoph and Bergamasco, Enrico and Grüninger, Markus and Bordovalos, Alexander V. and Chaulagain, Suresh and Podraza, Nikolas J. and Tokarz, Waldemar and Tabis, Wojciech and Wahila, Matthew J. and Sarker, Suchismita and Pollock, Christopher J. and Shang, Shun-Li and Liu, Zi-Kui and Artrith, Nongnuch and de Groot, Frank M. F. and Benedek, Nicole A. and Singer, Andrej and Muller, David A. and Schlom, Darrell G.},
    title = {{Synthesis of epitaxial \ch{TaO2} thin films on \ch{Al2O3} by suboxide molecular-beam epitaxy and thermal laser epitaxy}},
    journal = {APL Materials},
    volume = {14},
    number = {7},
    pages = {071106},
    year = {2026},
    month = {07},
    issn = {2166-532X},
    doi = {10.1063/5.0322051},
    url = {https://doi.org/10.1063/5.0322051},
}

@article{zhao2017lead,
  title={{Lead-free antiferroelectric silver niobate tantalate with high energy storage performance}},
  author={Zhao, Lei and Liu, Qing and Gao, Jing and Zhang, Shujun and Li, Jing-Feng},
  journal={Adv. Mater.},
  volume={29},
  number={31},
  pages={1701824},
  year={2017},
  publisher={Wiley Online Library}
}

@article{liu2018antiferroelectrics,
  title={Antiferroelectrics for energy storage applications: a review},
  author={Liu, Zhen and Lu, Teng and Ye, Jiaming and Wang, Genshui and Dong, Xianlin and Withers, Ray and Liu, Yun},
  journal={Adv. Mater. Technol.},
  volume={3},
  number={9},
  pages={1800111},
  year={2018},
  publisher={Wiley Online Library}
}

@article{soon2010dielectric,
  title={{Dielectric and soft-mode behaviors of \ch{AgTaO3}}},
  author={Soon, Hwee Ping and Taniguchi, Hiroki and Itoh, Mitsuru},
  journal={Phys. Rev. B},
  volume={81},
  number={10},
  pages={104105},
  year={2010},
  publisher={APS}
}

@article{li2025high,
  title={{A high-resolution neutron diffraction and electron diffraction study of the structure and phase transitions in \ch{AgTaO3}}},
  author={Li, JWB and Mullens, Bryce G and Torii, Shuki and Ji, Wenhai and Tan, Zhenhong and Xie, Wu and Kamiyama, Takashi and Miao, Ping and Kennedy, Brendan J},
  journal={{APL Materials}},
  volume={13},
  number={9},
  year={2025},
  publisher={AIP Publishing}
}

@article{francombe1958structural,
  title={{Structural and electrical properties of silver niobate and silver tantalate}},
  author={Francombe, MH and Lewis, B},
  journal={Acta Crystallogr.},
  volume={11},
  number={3},
  pages={175--178},
  year={1958},
  publisher={International Union of Crystallography}
}

@misc{belyaev1978orthorhombic,
  title={{Orthorhombic silver metatantalate and solid-solutions of system \ch{(Ag,Na)TaO3}, \ch{Ag(Nb,Ta)O3} }},
  author={Belyaev, IN and Lupeiko, TG and Nalbandyan, VB},
  journal={Kristallografiya},
  volume={23},
  number={3},
  pages={620--621},
  year={1978},
  publisher={MEZHDUNARODNAYA KNIGA 39 DIMITROVA UL., 113095 MOSCOW, RUSSIA}
}

@article{suchanicz2009uniaxial,
  title={Uniaxial Pressure Effect on Dielectric Properties of AgTaO3 Single Crystals},
  author={Suchanicz, Jan and Kania, Antoni},
  journal={Ferroelectrics},
  volume={393},
  number={1},
  pages={21--26},
  year={2009},
  publisher={Taylor \& Francis}
}

@article{khan2021structural,
  title={{Structural, electronic, optical and thermoelectric properties in the phases of \ch{AgTaO3}}},
  author={Khan, S. A. and  Khan, H. U. and Mehmood  S. and Ali, Z.},
  journal={Mater. Sci. Semicond. Process.},
  volume={122},
  pages={105467},
  year={2021},
  publisher={Elsevier}
}

@article{PhysRevLett.80.1988,
  title = {{Effect of Mechanical Boundary Conditions on Phase Diagrams of Epitaxial Ferroelectric Thin Films}},
  author = {Pertsev, N. A. and Zembilgotov, A. G. and Tagantsev, A. K.},
  journal = {Phys. Rev. Lett.},
  volume = {80},
  issue = {9},
  pages = {1988--1991},
  numpages = {0},
  year = {1998},
  month = {Mar},
  publisher = {American Physical Society}
}

@article{PhysRevB.69.212101,
  title = {{Ab initio study of the phase diagram of epitaxial \ch{BaTiO3}}},
  author = {Di\'eguez, Oswaldo and Tinte, Silvia and Antons, A. and Bungaro, Claudia and Neaton, J. B. and Rabe, Karin M. and Vanderbilt, David},
  journal = {Phys. Rev. B},
  volume = {69},
  issue = {21},
  pages = {212101},
  numpages = {4},
  year = {2004},
  month = {Jun},
  publisher = {American Physical Society}
}

@article{schwaigert2023molecular,
  title={{Molecular beam epitaxy of \ch{KTaO3}}},
  author={Schwaigert, Tobias and Salmani-Rezaie, Salva and Barone, Matthew R and Paik, Hanjong and Ray, Ethan and Williams, Michael D and Muller, David A and Schlom, Darrell G and Ahadi, Kaveh},
  journal={J. Vac. Sci. Technol., A},
  volume={41},
  number={2},
  year={2023},
  publisher={AIP Publishing}
}

@article{gupta2022ktao3,
  title={{\ch{KTaO3}-the new kid on the spintronics block}},
  author={Gupta, Anshu and Silotia, Harsha and Kumari, Anamika and Dumen, Manish and Goyal, Saveena and Tomar, Ruchi and Wadehra, Neha and Ayyub, Pushan and Chakraverty, Suvankar},
  journal={Adv. Mater.},
  volume={34},
  number={9},
  pages={2106481},
  year={2022},
  publisher={Wiley Online Library}
}

@article{schwaigert2026synthesis,
  title={{Synthesis and Electronic Structure of the Fractionally Occupied Double Perovskite \ch{EuTa2O6} with Ordered Europium Vacancies}},
  author={Schwaigert, Tobias and Barooni, Ali and Gregory, Benjamin and Malinowski, Paul and Tenneti, Anirudh and Hasko, Sonia and Palazzolo, Brenan and Hodgson, Jeffrey W and Faeth, Brendan and Woodward, Patrick M and Shen, K. M. and Singer, A. and Ghazisaeidi, M and Salmani-Rezaie, Salva and Schlom, G. S. and Ahadi, K.},
  journal={Adv. Funct. Mater.},
  volume={36},
  number={10},
  pages={e13656},
  year={2026},
  publisher={Wiley Online Library}
}

@article{schlom2007strain,
  title={{Strain tuning of ferroelectric thin films}},
  author={Schlom, Darrell G and Chen, Long-Qing and Eom, Chang-Beom and Rabe, Karin M and Streiffer, Stephen K and Triscone, Jean-Marc},
  journal={Annu. Rev. Mater. Res.},
  volume={37},
  number={1},
  pages={589--626},
  year={2007},
  publisher={Annual Reviews}
}

@article{arnault2023anisotropic,
  title={{Anisotropic superconductivity at \ch{KTaO3} (111) interfaces}},
  author={Arnault, Ethan G and Al-Tawhid, Athby H and Salmani-Rezaie, Salva and Muller, David A and Kumah, Divine P and Bahramy, Mohammad S and Finkelstein, Gleb and Ahadi, Kaveh},
  journal={Science Advances},
  volume={9},
  number={7},
  pages={eadf1414},
  year={2023},
  publisher={American Association for the Advancement of Science}
}

@article{al2025spin,
  title={{Spin-to-charge conversion at \ch{KTaO3} (111) interfaces}},
  author={Al-Tawhid, Athby H and Sun, Rui and Comstock, Andrew H and Kumah, Divine P and Sun, Dali and Ahadi, Kaveh},
  journal={Appl. Phys. Lett.},
  volume={126},
  number={9},
  year={2025},
  publisher={AIP Publishing}
}

@article{poage2025violation,
  title={Violation of the Pauli limit at KTaO 3 (110) interfaces},
  author={Poage, Samuel J and Gao, Xueshi and Baksi, Merve and Salmani-Rezaie, Salva and Muller, David A and Kumah, Divine P and Lau, Chun Ning and Lorenzana, Jos{\'e} and Gastiasoro, Maria N and Ahadi, Kaveh},
  journal={Phys. Rev. B},
  volume={111},
  number={21},
  pages={214506},
  year={2025},
  publisher={APS}
}

@article{al2023enhanced,
  title={{Enhanced critical field of superconductivity at an oxide interface}},
  author={Al-Tawhid, Athby H and Poage, Samuel J and Salmani-Rezaie, Salva and Gonzalez, Antonio and Chikara, Shalinee and Muller, David A and Kumah, Divine P and Gastiasoro, Maria N and Lorenzana, Jos{\'e} and Ahadi, Kaveh},
  journal={Nano Letters},
  volume={23},
  number={15},
  pages={6944--6950},
  year={2023},
  publisher={ACS Publications}
}

@article{mccourt2025electrostatic,
  title={{Electrostatic control of quantum phases in \ch{KTaO3}-based planar constrictions}},
  author={McCourt, Jordan T and Arnault, Ethan G and Baksi, Merve and Poage, Samuel J and Salmani-Rezaie, Salva and Ahadi, Kaveh and Kumah, Divine and Finkelstein, Gleb},
  journal={Nano Letters},
  volume={25},
  number={45},
  pages={16091--16096},
  year={2025},
  publisher={ACS Publications}
}

@article{vicente2021spin,
  title={{Spin--charge interconversion in \ch{KTaO3} 2D electron gases}},
  author={Vicente-Arche, Luis M. and Bréhin, Julien and Varotto, Sara and Cosset-Cheneau, Maxen and Mallik, Srijani and Salazar, Raphaël and Noël, Paul and Vaz, Diogo C. and Trier, Felix and Bhattacharya, Suvam and Sander, Anke and Le Fèvre, Patrick and Bertran, François and Saiz, Guilhem and Ménard, Gerbold and Bergeal, Nicolas and Barthélémy, Agnès and Li, Hai and Lin, Chia-Ching and Nikonov, Dmitri E. and Young, Ian A. and Rault, Julien E. and Vila, Laurent and Attané, Jean-Philippe and Bibes, Manuel},
  journal={Adv. Mater.},
  volume={33},
  number={43},
  pages={2102102},
  year={2021},
  publisher={Wiley Online Library}
}

@article{kim2024electronic,
  title={{Electronic-grade epitaxial (111) \ch{KTaO3} heterostructures}},
  author={Jieun Kim  and Muqing Yu  and Jung-Woo Lee  and Shun-Li Shang  and Gi-Yeop Kim  and Pratap Pal  and Jinsol Seo  and Neil Campbell  and Kitae Eom  and Ranjani Ramachandran  and Mark S. Rzchowski  and Sang Ho Oh  and Si-Young Choi  and Zi-Kui Liu  and Jeremy Levy  and Chang-Beom Eom },
  journal={Sci. Adv.},
  volume={10},
  number={21},
  pages={eadk4288},
  year={2024},
  publisher={American Association for the Advancement of Science}
}

@article{zou2015latio3,
  title={{\ch{LaTiO3}/\ch{KTaO3} interfaces: A new two-dimensional electron gas system}},
  author={Zou, K and Ismail-Beigi, Sohrab and Kisslinger, Kim and Shen, Xuan and Su, Dong and Walker, FJ and Ahn, CH},
  journal={APL Materials},
  volume={3},
  number={3},
  year={2015},
  publisher={AIP Publishing}
}

@article{kaspar2010spectroscopic,
  title={{Spectroscopic evidence for Ag (III) in highly oxidized silver films by X-ray photoelectron spectroscopy}},
  author={Kaspar, Tiffany C and Droubay, Tim and Chambers, Scott A and Bagus, Paul S},
  journal={J. Phys. Chem. C},
  volume={114},
  number={49},
  pages={21562--21571},
  year={2010},
  publisher={ACS Publications}
}

@article{KHANUJA200941,
title = {{XPS depth-profile of the suboxide distribution at the native oxide/Ta interface}},
journal = {Journal of Electron Spectroscopy and Related Phenomena},
volume = {169},
number = {1},
pages = {41-45},
year = {2009},
issn = {0368-2048},
doi = {https://doi.org/10.1016/j.elspec.2008.10.004},
author = {Manika Khanuja and Himani Sharma and B.R. Mehta and S.M. Shivaprasad},
}

@article{friedrich1913,
  title={Interferenzerscheinungen bei roentgenstrahlen},
  author={Friedrich, Walther and Knipping, Paul and Laue, Max},
  journal={Annalen der Physik},
  volume={346},
  number={10},
  pages={971--988},
  year={1913},
  publisher={Wiley Online Library}
}

@article{nelson1945,
  title={An experimental investigation of extrapolation methods in the derivation of accurate unit-cell dimensions of crystals},
  author={Nelson, Jo Bo and Riley, Dennis P},
  journal={Proceedings of the Physical Society},
  volume={57},
  number={3},
  pages={160--177},
  year={1945}
}

\end{document}